\RequirePackage{fix-cm}
\documentclass[twocolumn]{svjour3}          % twocolumn
\smartqed  % flush right qed marks, e.g. at end of proof

\usepackage[utf8]{inputenc}
\usepackage{graphicx} % for pdf, bitmapped graphics files
\usepackage{mathptmx} % assumes new font selection scheme installed
\usepackage{amsmath} % assumes amsmath package installed
\usepackage{color}

\usepackage{color} 
\usepackage{float}
\usepackage{bm}
\usepackage{url}
\usepackage{tabularx}
\usepackage{arydshln}
\usepackage{amsfonts}

\usepackage{siunitx}
\usepackage{booktabs} % <-- To get prettier rules in tables
\usepackage{caption} % <-- To set caption width etc.
\journalname{Preprint}

\AtBeginDocument{%  
  \setlength{\oddsidemargin}{\dimexpr(\paperwidth-\textwidth)/2-1in}%
  \setlength{\evensidemargin}{\oddsidemargin}%
  \setlength{\topmargin}{%
    \dimexpr(\paperheight-\textheight)/2-\headheight-\headsep-1in}%
}

\begin{document}

\title{Vibration Analysis in Bearings for Failure Prevention using CNN}
%\title{Vibration Analysis with Images for Bearings Wear Classification using CNNs

%\thanks{*This work was supported by the Mexican National Council of Science and Technology CONACYT, and the FORDECyT project 296737 “Consorcio en Inteligencia Artificial”.}



\author{L.A. Pinedo-Sánchez$^{1}$ \and D.A. 
        Mercado-Ravell$^{1,2}$ \and
        C.A. Carballo-Monsivais$^{1}$ %etc.
}


\institute{$^{1}$Authors are with the Center for Research in Mathematics CIMAT AC, campus Zacatecas. \at
              Avenida Lasec Andador Galileo Galilei, Manzana 3 Lote 7,
              Parque Quantum, ZACATECAS 98160
              Mexico\\
              Tel.: 492-998-0300\\
              \email{luis.pinedo@cimat.mx, 
              abraham@cimat.mx} %  \\
           \and
           $^{2}$ Is also with Cátedras CONACYT, at CIMAT-Zacatecas \\
           \email{diego.mercado@cimat.mx}
}

\date{May 6th 2020}

\maketitle
\begin{abstract}
The timely failure detection for bearings is of great importance to prevent economic loses in the industry. In this article we propose a method based on Convolutional Neural Networks (CNN) to estimate the level of wear in bearings. First of all, an automatic labeling of the raw vibration data is performed to obtain different levels of bearing wear, by means of the Root Mean Square features along with the Shannon's entropy to extract features from the raw data, which is then grouped in seven different classes using the K-means algorithm to obtain the labels. Then, the raw vibration data is converted into small square images, each sample of the data representing one pixel of the image. Following this, we propose a CNN model based on the AlexNet architecture to classify  the wear level and diagnose the rotatory system. To train the network and validate our proposal, we use a dataset from the center of Intelligent Maintenance Systems (IMS), and extensively compare it with other methods reported in the literature. The effectiveness of the proposed strategy proved to be excellent, outperforming other approaches in the state-of-the-art.

\keywords{ Vibration analysis \and Bearing fault \and Deep learning \and Image classification \and CNN \and AlexNet }

\end{abstract}

\section{Introduction}
\label{Introduction}

Rotating mechanisms are essential components in most industrial machines, and the problems that can occur in rotating machines cause significant losses to the industry every year. One of the key components of rotating machines are the bearings, and these are exposed to excessive wear due to the many hours of continuous operation. Within the industry, rotary machinery is used in numerous forms including pumps, electric motors, power generators, ventilators, wind turbines, alternative compressors, refrigeration towers, among others.

Signal processing has helped organizations operating machinery to prevent failures, along with other kind of problems, such as low productivity, safety risks, downtime, among others \cite{Mahamad2010,Xia2018}. Henceforth, it is fundamental that the detection of machinery failures is accomplished on time, to avoid problems in the future and to improve the performance of the organizations, helping to prevent stopping the production, causing permanent damage on expensive components, complete machine failure or even an accident. 

Moreover, rotating machines always incorporate bearings, and these components are substantial in their functioning \cite{Qiu2006}. The bearings are exposed to wear, causing the machines not to operate in favorable conditions and lose efficiency. As the wear on the bearing is higher, the vibration signals increase compromising the system's performance, however, such vibrations can also be exploited to detect failures without stopping the production, generation important savings to the companies. Furthermore, analyzing vibrations on the bearings can also be utilized to detect problems with other components of the rotating system. For this reason, the analysis of bearing vibrations is of great importance for failure detection and monitoring of the machine health condition \cite{Xie2017}.

%In recent years, researchers have studied bearing condition monitoring and failure diagnosis. These studies have focused on feature extraction and condition recognition \cite{Tabrizi2017}.

%parrafo para la tesis
Different techniques have been used to monitor rotating machines. For example, in \cite{BenAli2015}, the authors proposed a method that combines a Simplified Fuzzy Adaptive Resonance Theory Map (SFAM) neural network and Weibull distribution. In \cite{Xuan2017}, the Hilbert-Huang transform is used to obtain the frequency energy values of the vibration signals of a motor, which is analyzed through a Support Vector Machine (SVM) for prediction failure. Also, in \cite{Lu2020} a method making use of the Complementary Ensemble Empirical Mode Decomposition (CEEMD) is presented, with a kernel of SVM to make the evaluation of the health condition of the bearings. As well as in \cite{Akuruyejo2018}, where they proposed a method to estimate the Remaining Useful Life (RUL) of bearings, where the Decomposition Empirical Mode method is used, along with the Principal Component Analysis and Support Vector Regression (SVG) algorithms. In addition, in \cite{Mortada2011} they applied an approach called Logical Data Analysis, to perform failure detection on rotating machinery using vibration signals. Furthermore in \cite{Zhang2017a} they proposed a method based on Deep Neural Networks (DNN), where they created models with different numbers of layers and perform the recognition of the type of fault that occurs. While in \cite{Soualhi2014} it is used a technique called Artificial Ant Clustering to detect the degradation state of the bearings, the hidden Markov models are utilized to give an approximation of the next degradation state and an adaptive system of neurodiffuse inference, along with time series predictions, are applied to make the estimation of the remaining time to the next degradation state. 

The main problem of the classical techniques is that they require the supervision of an expert in the area for the extraction of characteristics, thus, more advanced engineering is required, which involves greater human effort \cite{Wen2018}. Some of these methods are still used today to perform fault classification, such as SVM. However, these methods have to be combined with other more recent techniques, and considerably modified to improve their performance \cite{Lu2020,Hemmer2018}. 

More recently, neural networks have successfully been carried out for this task. Some of the first attempts include \cite{Mahamad2010}, where it is proposed an Artificial Neural Network (ANN) to make the accurate prediction of the RUL of the bearings. The ANN they used was a Feed Forward Neural Networks with a Levenberg Marquardt's training algorithm. Also, an Elman Neuronal Network (ENN) was proposed in \cite{Kramti2018} to predict the RUL in wind turbine generators, where the ENN output is the percentage of RUL. However, these neural network based techniques have been considerably overcome by modern Deep Learning approaches.

Currently, there are new techniques and tools such as Deep Learning, which allow the extraction of features automatically and without the need for an expert, simplifying the final solution while considerably improving the accuracy \cite{Raghu2017}. Furthermore, this kind of methodologies can be easily generalized or transferred to another different context \cite{Zhang2019a}.

It is then of great importance to take advantage of these new techniques, such as Deep Learning, as it can be of great help in predicting bearing failures. Henceforth, in recent years, researchers have studied bearing condition monitoring and failure diagnosis using Deep Learning. These studies have focused on feature extraction and condition recognition \cite{Tabrizi2017}, where the use of Convolutional Neural Networks (CNN) have proven to be an excellent option for effectively learning features in the context of fault diagnosis through vibration analysis \cite{Xia2018,Wen2018,Hemmer2018,Li2017,Zilong2018,Zhang2018,Guo2016,Xu2019,Wang2019}.

This work focuses on the development of an automatic method to prevent failure in rotary machines by wear estimation in bearings, where the intervention of an expert is not required. Accordingly, we propose a method for vibration analysis in bearings through images using CNN, where the information is obtained by means of accelerometers, which is advantageous since it does not require to stop the production periodically to check the machine deterioration. Besides, having a constant monitoring of the machines automatically provides a way to give maintenance in a timely way. In order to make use of supervised learning and take advantage of the great advances in image processing with CNNs, we propose a strategy to perform labeling of raw vibration signals, which are then transformed to small square images for training and classification. The proposed labeling strategy consists in the extraction of the Root Mean Square (RMS) along with the Shannon's entropy of the raw signal, which is then grouped in seven different categories using the K-means algorithm to obtain the classes for labeling. Afterwards, sub-samples of the raw signals are converted into images for training a CNN in order to automatically extract characteristics and create a model to classify the wear of the bearings over time. In this work, the AlexNet architecture is taken, where some modifications were made to adapt it to the problem under consideration. The proposed methodology was implemented and studied using the Intelligent Maintenance Systems (IMS) unlabeled dataset \cite{NationalAeronauticsandSpaceAdministration2018}. Last but not least, extensive comparison with other classical techniques, as well as with state-of-the-art CNN based methods revealed that the proposed strategy considerably outperforms other approaches in the literature for the IMS dataset.

This article is organized in the following way: firstly, Section \ref{Related Works} introduces the related works on bearing failure detection using CNN. Afterwards, Section \ref{Methodology} describes the proposed methodology for estimating the level of wear in bearings, including the proposed CNN architecture. Later on, Section \ref{Experiments and results} validates the proposal and shows the obtained results, including and extensive comparison with other available methods. Finally, Section \ref{Conclusion} presents the conclusions and future work.

\section{Related Works}
\label{Related Works}
\begin{table*}[th!] %layerLeNet
\centering
 \caption{Comparison between the main bearing failure detection works.}
 \begin{tabular}{c c c c c c c}  
 \hline \\
 Work & Based on/Architecture &  Dataset & \begin{tabular}[c]{@{}l@{}}Converts \\to image\end{tabular} & Domain & Goal & Accuracy \vspace{0.2cm} \\ 
 \hline \\
 \cite{BenAli2015} & \begin{tabular}[c]{@{}l@{}}Simplified Fuzzy Adaptive \\ Resonance Theory Map \\ and Weibull distribution\end{tabular} & IMS & No & Time & RUL & $65.46\%$  \\
 \cite{Xuan2017} & Gray model & Own & No & Frequency & Failure detection & NA \\
 \cite{Lu2020} & SVM & IMS & No & Frequency & Wear level & $98.50\%$  \\
 \cite{Akuruyejo2018} & SVG & IMS & No & Frequency & RUL & NA \\
 \cite{Zhang2017a} & Deep Neural Networks & CWRU and IMS & No & Time & Failure detection & $98.35\%$  \\
 \cite{Li2017} & CNN & CWRU & Yes &  Frequency & Failure detection & $97.89\%$  \\
 \cite{Xia2018} & CNN & CWRU and gearbox & Yes & \begin{tabular}[c]{@{}l@{}}Time and \\Frequency\end{tabular} & Failure detection & $98.35\%$  \\
 \cite{Zilong2018} & CNN & CWRU & Yes & Time & Failure detection & $98.69\%$  \\
 \cite{Zhang2018} & CNN & CWRU & Yes & Time & Failure detection & $99.22\%$ \\
 \cite{Eren2019} & CNN & CWRU and IMS & No & Time & Failure detection & $93.90\%$  \\
 \cite{Guo2019} & CNN & \begin{tabular}[c]{@{}l@{}} CWRU, IMS and Railway\\ Locomotive (RL)\end{tabular} & No & Time & Failure detection & $86.30\%$  \\
 \cite{Guo2016} & CNN/LeNet-5 & CWRU & Yes & Time & Failure detection & $97.90\%$  \\
 \cite{Wen2018} & CNN/LeNet-5 & CWRU and pumps & Yes & Time & Failure detection & $97.04\%$  \\
 \cite{Xu2019} & CNN/LeNet-5 & CWRU & Yes & Time & Failure detection & $99.40\%$ \\
 \cite{Hemmer2018} & \begin{tabular}[c]{@{}l@{}}CNN/AlexNet, SVM \\and SAE-SVM\end{tabular} & \begin{tabular}[c]{@{}l@{}}Radial Roller Bearing\\ Test Rig and Axial\\ Roller Bearing Test Rig\end{tabular} & Yes & Frequency & Failure detection & $91.96\%$  \\
 \cite{Wang2019} & CNN/AlexNet & CWRU & Yes & Frequency & Failure detection & $94.39\% - 100\%$  \\
 Proposal & CNN/AlexNet & IMS & Yes & Time & Wear level &  $99.25\%$\\ [1ex] 
 \hline
 \end{tabular}
\label{table:RelatedWorks}
\end{table*}

In the following, we discuss the most relevant recent works regarding the diagnosis of bearing failures by means of vibrations. More in particular, this Section reviews modern CNN based techniques applied to this particular task, as well as the most important architectures up to date, such as LeNet-5 and AlexNet. The most relevant works are shown in Table \ref{table:RelatedWorks}, where the works performed with classical and CNN-based methods are compared. 

%\textcolor{red}{agregar parrafo explicando la idea de la conversi[on de se;ales a imagenes y que esta a permitido explotar las tecnicas de CNN desarrolladas para clasificaci[on de imagenes, que es una importante alternativa para usar CNN }

With the recent huge advances within the CNN, a large number of applications have been made to solve the classification, detection and segmentation problems, particularly in the fields of computer vision and image processing. These applications have had very good results, due to the potential that CNN have to extract a large number of characteristics, and generalize to different scenarios, producing great impact improving previous results \cite{Wen2018,ZHANG2019}. More recently, work has been developed for the detection of bearing failures by making use of CNN, where the data from the vibration signals is transformed to images, taking advantage of the enormous advances obtained in image processing with CNN, hence, becoming an excellent strategy for vibration analysis.

%\textcolor{red}{sugerencia: y si pones una tabla con los principales trabajos relacionados donde hagas una comparacion cualitativa, diceindo si es basado en CNN, tipo de CNN, tipo de DL, tipo de arquitectura, alexnet, lanet, que basee de datos usa, que extractor usa c/u, que tipo de clasificacion, si convierte a imagen, y la precision obtenida, si hacen conversion a imagenes, tiempo o frecuencia, etc.}

\subsection{Convolutional Neural Network}

%\textcolor{red}{aqui empezaria yo la seccionn de trabajos relacionados. La introduccion conncluiria con que lo de hoy son las CNN}

With the advances that have emerged in the field of deep learning, several methods have been developed to diagnose bearings failures, with one of the most popular models being the CNN \cite{Wang2019}. CNN are methods that have a great ability to learn in the areas of image classification, object detection, text recognition, etc. \cite{Pan2018}, reason why this kind of methods have become very important for digital signal processing \cite{Zilong2018}.

Recently, CNN based methods have been implemented to perform bearing failure prediction, as in \cite{Li2017} where the authors propose a method based on CNN in combination with the improved Dempster-Shafer theory called IDSCNN. There, the CNN architecture consists of three Convolutional Layers (Conv) and two Fully Connected (FC) layers. Also, the RMS of the Fast Fourier Transform is utilized. In their experiments they used the Case Western Reserve University (CWRU) dataset for validations. In addition, in \cite{Xia2018} a method based on CNN is proposed to make the diagnosis of failures in rotating machines, where they make use of multiple accelerometers and combine the obtained information to create a two-dimensional matrix. Also, in \cite{Zilong2018} they propose a multi-scale convolution method called MS-DCNN where they reduce the number of parameters and training time. This method is compared with one and two dimensional CNN. Meanwhile, in \cite{Zhang2018} a method was proposed where they make the classification of vibrations for the diagnosis of failures, detecting the source of failure and the different degrees of damage. The vibration signal was converted to a spectrogram by means of the Short-Time Fourier Transform (STFT), which was used as input of a CNN to make the training of the data. They classified the vibrations into seven states according to the ring of the bearing (inner or outer), and its wearing level. On the other side, in \cite{Eren2019} they propose a one-dimensional CNN, where only seven layers are used to detect the type of failure in the bearings. Also, in \cite{Guo2019} they propose a method called Deep Convolutional Transfer Learning Network (DCTLN) that has two main modules: condition recognition and domain adaptation. The first module is a one dimensional CNN that is in charge of learning the characteristics of the data and recognizing the condition of the machines. The second module helps the first module to learn the characteristics of data variations. In all the before mentioned works, the employed datasets are already labeled according to wear level and type of failure, which is a great advantage when you want to perform supervised learning. Unfortunately, not all the available datasets are properly labeled, and it is not clear how to classify different data to apply the aforementioned techniques. Furthermore, the available labels only consider whether or not there is a failure, but do not provide the level of wear, which may be useful to prevent failures before they occur by opportune maintenance.

%\textcolor{red}{nuevvamente, no hay conexion entre los trabajos que abordas, falta fluidez y discusion, trata de agrupar los que son similares. Tambien resalta sus inconvenientes y contrastalos con tu trabajo. Cada vez te vas acerrcando mas a lo que tu haces y hay que ir diferrenciando con lo tuyo para justificar porque tu trabajo es reelevane}

At current state-of-the-art, two main CNN architectures have been used as a base to obtain the best results reported in the literature for classification of bearing failure using images, LeNet-5 \cite{Lecun1998} and AlexNet \cite{Krizhevsky1012}. A review on their main adaptations for this particular task is presented in the following.

\subsubsection{LeNet-5}

LeNet-5 is a classic CNN architecture for performing image classification, such as handwritten or machine printed character recognition, and multi-object detection  \cite{Ren2017,Wen2018,Hong2016}.

This architecture was proposed by \cite{LeCun1990,Lecun1998}, and includes seven layers, not counting the input layer. Two convolutional layers (Conv), two sub-sampling layers and three FC layers. The configuration of the layers is shown in Table \ref{table:layersLeNet}. For the CNN input, the usual resolution is $32 \times 32$ pixels. However, these input dimensions may vary, depending on the data size.

\begin{table}[hb!] %layerLeNet
\centering
 \caption{LeNet-5 layers configuration.}
 \begin{tabular}{c c}  
 \hline \\
 Layer & Features \vspace{0.2cm} \\ 
 \hline \\
 1 &  Conv(5 x 5 x 6)  \\
 2 &  Avgpool(2 x 2) \\
 3 &  Conv(5 x 5 x 16)  \\
 4 &  Avgpool(2 x 2)  \\
 5 &  FC(120)  \\
 6 &  FC(84)  \\
 7 &  FC(10)  \\  [1ex] 
 \hline
 \end{tabular}
\label{table:layersLeNet}
\end{table}

Several CNN architectures based on LeNet-5 have been used in the literature, under different modifications in order to improve the results, as in \cite{Guo2016} where it is proposed a method called ADCNN inspired by the LeNet-5 architecture, but adding a convolutional and a sub-sampling layers before the first FC one. With the proposed method, the classification of the bearing failure type is made, along with the severity of the failure. Also, this work makes use of the CWRU labeled dataset. Furthermore, an architecture based on LeNet-5 was proposed in \cite{Wen2018}, where two more convolution and two grouping layers are added. The transformation of the images was performed in time-domain, using three datasets including the CWRU bearing dataset, a self-priming centrifugal pump dataset, and an axial piston hydraulic pump dataset. Also, in \cite{Xu2019} was proposed a method based on CNN and Random Forest (RF). Time-domain vibration signals containing fault information by Continuous Wavelet Transform (CWT) were transformed into images. The proposed method is based on LeNet-5 and the features extracted by the CNN are used by multiple RF classifiers to make the diagnosis of bearing failures.

The modification proposals of the this architecture have been designed to improve the classification of bearing failure based on images, reporting good results. The use of this architecture is adequate due to the size of the images being used, predominately small images. Unfortunately, in the literature where improvements to this architecture are performed, the authors do not normally present enough information about the configuration of the CNN to replicate the experiments, except for the work in \cite{Wen2018}.

\subsubsection{AlexNet}

AlexNet is a CNN architecture that was proposed by \cite{Krizhevsky1012}. This CNN has been satisfactorily used to classify images from the ImageNet LSVRC-2010 database, magnetic resonance images, traffic congestion detection, etc. \cite{Krizhevsky1012,Khagi2019,Ezhilarasi2018,Wang2018}. This architecture is not very deep, which is ideal for working with small images. Furthermore, it was responsible for the rise of CNN when used in the ImageNet dataset and good results were obtained \cite{Krizhevsky1012}.

This architecture is composed of eleven layers, not counting the input one, five convolutional layers, three sub- sampling layers and three FC layers. The configuration of the layers is shown in Table \ref{table:layersAlexNet}. Also, it is normally used for larger images than LeNet-5, around $224 \times 224$ pixels, since it employs more layers.

\begin{table}[hb!] %layerAlexNet
\centering
 \caption{AlexNet layers configuration.}
 \begin{tabular}{c c}  
 \hline \\
 Layer & Features \vspace{0.2cm}\\
 \hline \\
 1 &  Conv(11 x 11 x 96)  \\
 2 &  Maxpool(3 x 3) \\
 3 &  Conv(5 x 5 x 256)  \\
 4 &  Maxpool(3 x 3)  \\
 5 &  Conv(3 x 3 x 384)  \\
 6 &  Conv(3 x 3 x 384)  \\
 7 &  Conv(3 x 3 x 256) \\
 8 &  Maxpool(3 x 3) \\
 9 &  FC(4096) \\
 10 &  FC(4096) \\
 11 &  FC(1000) \\  [1ex] 
 \hline
 \end{tabular}
\label{table:layersAlexNet}
\end{table}

Only two different adaptations of AlexNet can be found in the literature for bearing failure diagnosis using images. For example, in \cite{Hemmer2018}, AlexNet was slightly modified and combined with two SVM classifiers and two Sparse Auto-Encoder SAE-SVM classifiers. The experiments were performed with frequency-domain images. Also, in \cite{Wang2019} was proposed a method for bearing failure diagnosis based on CNN with the AlexNet architecture. To validate their proposal it was necessary to convert the vibration signals into time frequency images, by using eight time frequency analysis methods. In their experiments the CWRU dataset was used. Although good results have been obtained using AlexNet for bearings diagnosis, the main differences with respect to the present work are that the datasets employed so far study bearings that were physically manipulated to provoke a failure. Therefore, these datasets already have labels corresponding to the location of the failure, but induced failures do not necessarily correspond to the normal deterioration under regular use. Moreover, they can not be tested with other datasets due to the lack of equivalent labels. Also, as can be observed in Table \ref{table:RelatedWorks}, most of the available works aim only to detect whether or not a failure is present, but do not provide the level of wear, which can be used to timely schedule a maintenance task to prevent failures. On the other hand, most of the related work make use of images in frequency-domain, while in the present work the images are made in time-domain. Furthermore, they do not make many modifications to the architecture, because they work with spectrogram images which are larger, resulting in higher training times. In addition, they present a combination of this architecture with another classification method, such as SVM. In the present work, modifications were made to the AlexNet architecture to adapt it to the size of the images, which are considerably smaller, significantly reducing the training time, while allowing to obtain a major number of images from the available data, which proved to be beneficial for the CNN classifier.

In this work, an automatic labeling strategy is proposed to classify the wear levels of the bearings over time, without the need of expert supervision, where the labels were made based on the similarity of characteristics of the data using Traditional Statistical Features (TSF) along with the Shannon's entropy. Seven types of classes were obtained by means of the K-means algorithm, and used for labeling of the raw vibration data, where one class is considered as a healthy state and the rest of the classes are the different levels of wear, which can be used to schedule the corresponding maintenance tasks and prevent failures. The vibration signals in time-domain are transformed into images to train the CNN, which provide us a better automatic characteristics extraction from the data. We propose an AlexNet based architecture, where the size of the convolution and sub-sampling filters are modified, and a sub-sampling layer is added, in order to deal with smaller images, which allows us to increase the number of images obtained, considerably improving the training stage. The proposed methodology was extensively evaluated using the IMS unlabeled dataset with excellent results, specially when compared with other works in the literature, such as CNN-based and classic methods.

The main contributions of this work are summarized in the following:

\begin{enumerate}
  \item A CNN based classifier is used to estimate bearing wear by means of vibration images, providing a diagnosis of the system without stopping production, which can be used to timely give maintenance and prevent failures in rotatory machines.
  \item A method for performing automatic vibration data labeling, without expert supervision, is introduced. This is accomplished by means of RMS along with Shannon's entropy and K-means for feature clustering.
  \item The AlexNet architecture was adapted and satisfactorily utilized for estimating the level of bearing wear. The size of the convolution and sub-sampling filters were modified to deal with smaller images, and a sub-sampling layer was added to improve the results.
  \item The proposed methodology was successfully tested with the IMS unlabeled dataset with excellent results, surpassing other methods reported in the literature.
\end{enumerate}

\section{Methodology}
\label{Methodology}

\begin{figure*}[th!] %methodology
\begin{center}
  \includegraphics[width=1\textwidth]{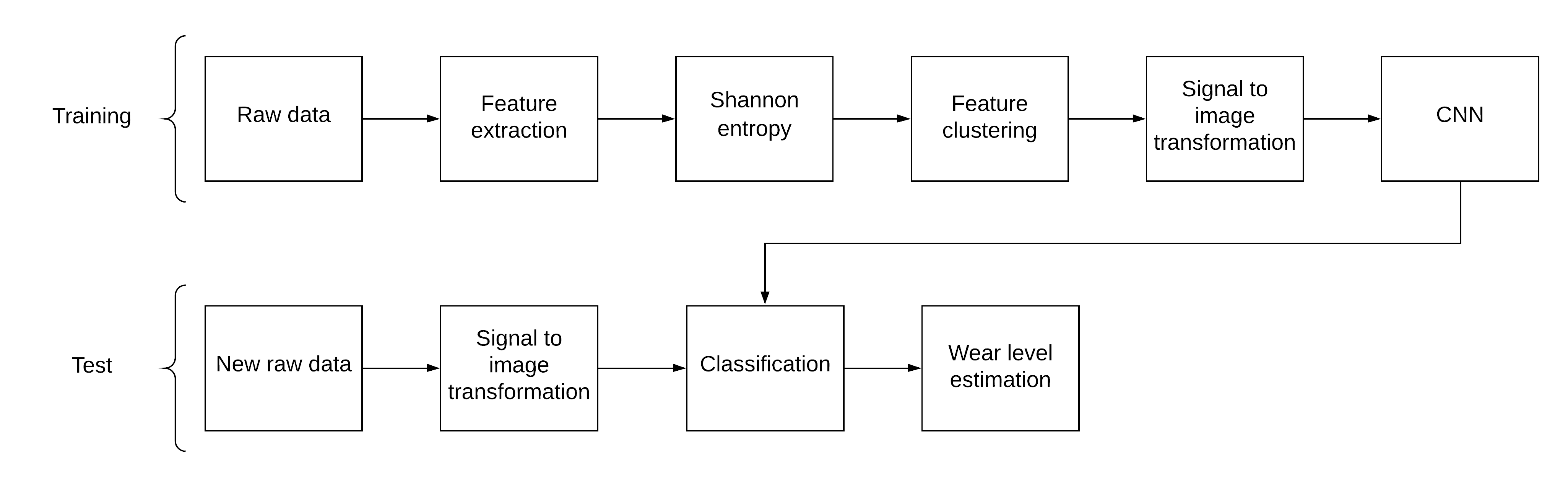}
\caption{Overall methodology for estimating the level of wear on bearings, where we perform the extraction of characteristics from the vibration signals, apply Shannon's entropy, afterwards make the clustering of characteristics to create classes, then make the transformation of the vibration signals to images, make the input of a CNN to create a model and perform the classification of the wear level of bearings.}
\label{fig:methodology}       % Give a unique label
\end{center}
\end{figure*}

We propose a methodology to estimate the level of wear in bearings by means of vibration signals in combination with CNN, where we perform a transformation of the raw signals to images. The Fig. \ref{fig:methodology} shows the steps that are followed to perform the estimation of the wear level in bearings. 

%\textcolor{red}{valdria la pena mencionar que se propone un nuevo etiquetaado automatico, inspirado en tecnicas clasicas, y citar el articulo del que te basaste}

A new labeling method for bearing vibration data, according to the level of wear is proposed, based on classical techniques \cite{BenAli2014}, where first the characteristics are extracted from the raw data using TSF, then Shannon's entropy is applied to highlight the extracted characteristics. Afterwards, the clustering of the characteristics is performed using the K-means algorithm to make the labeling of the data. This methodology is ideal when you have data that is not labeled by wear or damage level, since it provides an automatic way to classify it, providing the labels necessary for supervising learning, without intervention of a human expert. In this article we work with an unlabeled dataset provided by \cite{NationalAeronauticsandSpaceAdministration2018}, which is described in Section \ref{Experiments and results}.

Next, the conversion of the raw vibration signals to images is performed, and the images obtained are used to train a CNN for the classification task. We propose a CNN architecture based on AlexNet to make the wear level classifier. Finally, new raw data is obtained, and only a small section of the signal is required to be converted into images and the level of bearing wear is estimated.

\subsection{Feature extraction}

To perform the extraction of the characteristics from the vibration signals, it was necessary to make use of the TSF, because they are excellent tools that are used in time-domain to characterize the changes in the vibration signals of the bearings during operation. In addition, they also allow us to estimate the wear of the bearing over time. When bearings are damaged, the vibrations are intensified, and the TSF values increase considerably, indicating the damage. These characteristics are shown in Table \ref{table:TraditionalStatisticFeatures}, where $n$ indicates the number of discrete points in the sample, $x_{i}$ is a single experimental point from the sample, $\bar{x}$ is the mean of the sampled values, $\sigma$ represents the standard deviation and $min$ and $max$ are the minimum and maximum values of the sample, respectively.

\begin{table}[hb!] %Traditional Statistical Features
\centering
 \caption{Traditional Statistical Features (TSF).}
\label{table:TraditionalStatisticFeatures}
 \begin{tabular}{c c}  
 \hline \\
 Name & Formula \vspace{0.2cm} \\ 
 \hline \\
 RMS  & \large{ $\sqrt{\left ( \frac{1}{n} \sum_{i = 1}^{n} x_{i}^{2} \right )}$} \vspace{0.2cm} \\ 
 \hline \\ 
 Kurtosis & \large{  $\frac{1}{n}\sum_{i = 1}^{n}\frac{\left ( x_{i} - \bar{x} \right )^{4}}{\sigma ^{4}}$ }\vspace{0.2cm} \\
 \hline \\
 Skewness &  \large{ $\frac{1}{n}\sum_{i = 1}^{n}\frac{\left ( x_{i} - \bar{x} \right )^{3}}{\sigma ^{3}}$} \vspace{0.2cm} \\
 \hline \\
 Peak to peak &  \large{ $x_{min} - x_{max}$ }\vspace{0.2cm} \\ 
 \hline \\
 Crest Factor &  \large{ $\frac{max\left | x_{i} \right |} {RMS}$} \vspace{0.2cm} \\ 
 \hline \\
 Shape Factor &  \large{ $\frac{RMS}{\frac{1}{n}\sum_{i = 1}^{n}\left | x_{i} \right |}$ }\vspace{0.2cm} \\
 \hline \\
 Impulse Factor &  \large{ $\frac{max\left | x_{i} \right |}{\frac{1}{n}\sum_{i = 1}^{n}\left | x_{i} \right |}$} \vspace{0.2cm} \\
 \hline \\
 Margin Factor &  \large{ $\frac{max\left | x_{i} \right |}{\left ( \frac{1}{n}\sum_{i = 1}^{n}\left | x_{i} \right |^{\frac{1}{2}} \right )^{2}}$ }\vspace{0.2cm} \\
 \hline
 \end{tabular}
\end{table}

In the literature related to prognosis and diagnosis of bearings failures, RMS and Kurtosis are the most widely used \cite{Mahamad2010,BenAli2015,Huang2007}. Meanwhile, Kurtosis is effective for detecting bearings failure at an early stage. Whereas, RMS represents the energy and power characteristics of vibration signals.

The main idea of these characteristics is to identify a monotonous trend, that is to say, when the bearing deteriorates, the value of these characteristics is increased or decreased to indicate the failure. When there is damage in the bearing and it is not detected by one of the TSF, it is, there is no significant increasing or decreasing in these features, TSF will not be of great help for the analysis, so you will have to choose another TSF.

\subsection{Shannon's entropy}

Shannon's entropy is the central part of information theory, and it is also known as the measure of uncertainty. Shannon's entropy $H(x)$ was introduced with communication theory in 1948 \cite{Kapur1992,Shannon1948}. Then, in \cite{BenAli2014} the original formula was modified, and  it is defined as
\begin{equation}
	H(x) = \frac{1}{n} \sum_{i = 1}^{n} - TSF(x_{i}) log_{2} TSF(x_{i})
	\label{Ec:Shannon}
\end{equation}
where $n$ is the length of the sliding window and $TSF$ represents the Traditional Statistical Feature that was selected in the previous stage.

Once obtained the TSF, the Shannon's entropy can be calculated, such that it allows us to highlight the characteristics obtained from the TSF. In this way, we can choose the TSF together with the Shannon's entropy to observe well defined increase or decrease of the data over time. Once we have identified one of the TSF together with the Shannon's entropy, we can move on to the clustering stage.

\subsection{Feature clustering}

The next step is to group the data obtained from Shannon's entropy along with one of the TSF, in order to obtain the labels needed to perform the CNN training. For this step, the K-Means algorithm is implemented to make the grouping of the data with similar characteristics, such that, it is possible to label the different levels of wear of the bearings. In this phase we have the possibility to choose the number of classes we want to label in our dataset.

The K-Means algorithm was proposed in \cite{Lloyd1982} and is one of the most important unsupervised classification algorithms that allows us to group data in a specific number of groups that have similar characteristics. These groups are called "clusters" and the number of clusters is defined by $K$. This algorithm consists of minimizing the sum of the euclidean distances of each of the points with respect to the centroid of the cluster. 

\subsection{Signal to image transformation}

Traditional methods for motor failure are based on statistical analysis, fuzzy logic expert systems or genetic systems. Extracting characteristics from raw signals is one of the main functions of these methods, since a good feature extraction has a great impact on the results \cite{Wen2018,ZHANG2019}. In contrast to traditional methods, we perform a data pre-processing method that converts raw vibration signals in time-domain to images, in order to take advantage of the powerful classification tools available for image processing using CNN \cite{Wen2018,Wang2019,Li2017,Guo2016,Xu2019,ZHANG2019,Do2011,Lu2017}. Moreover, converting the raw signals into images provides a good way to explore two-dimensional features \cite{Do2011}.

\begin{figure*}[th!] %signalToImage
\begin{center}
  \includegraphics[width=0.8\textwidth]{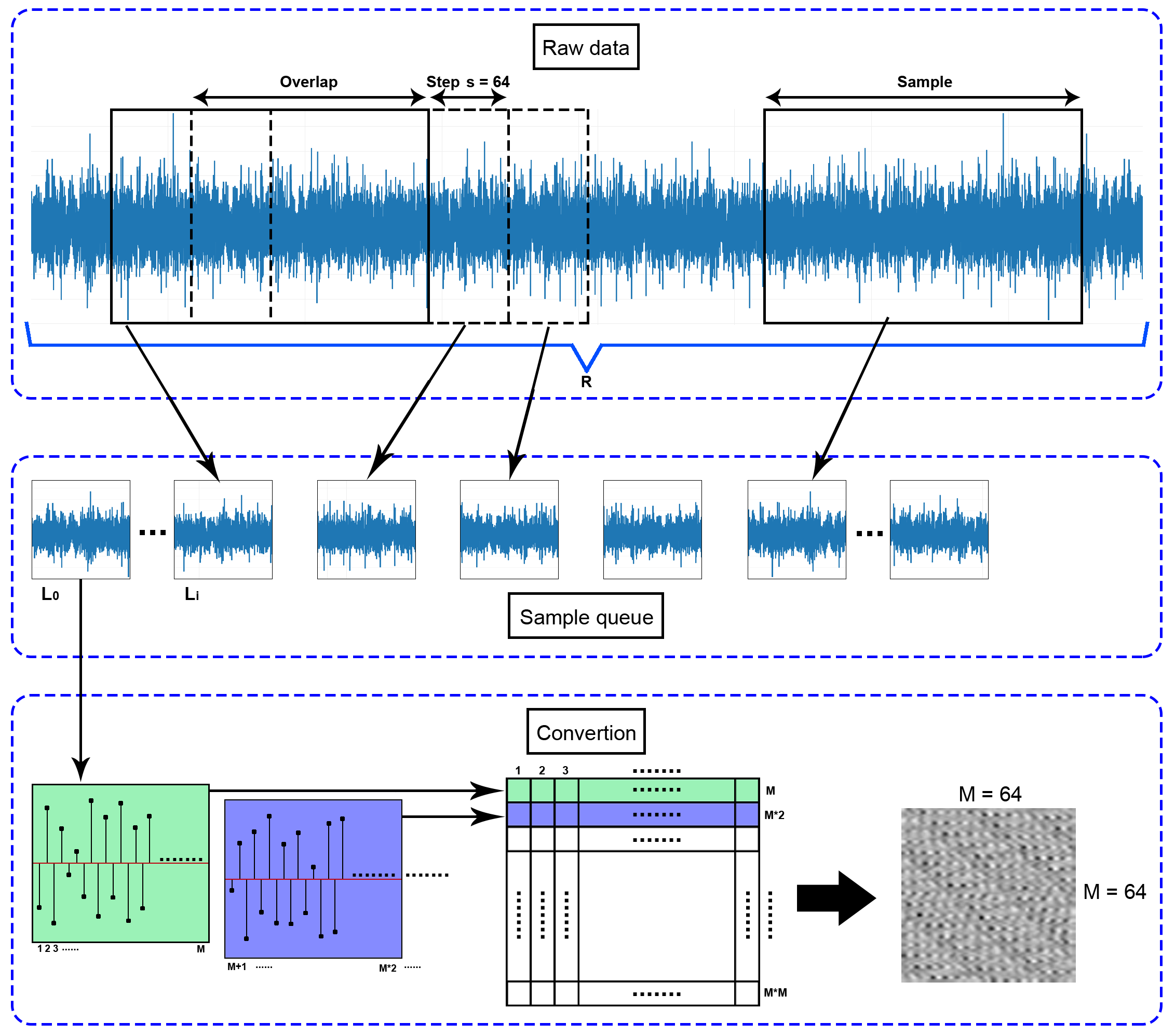}
\caption{Method to convert the raw signals into images. First, an $M ^2$ signal sub-sample is taken, where $M$ represents the total height and width of the square image. This sub-sample is then mapped into a matrix and each point is normalized in a range from $0$ to $255$ to represent the intensity of each pixel value.}
\label{fig:signalToImage}       % Give a unique label
\end{center}
\end{figure*}

For a raw signal $R$ with $N$ sample points, Fig. \ref{fig:signalToImage} shows the conversion method to images, where each time-domain signal point is one pixel of a square grayscale image with size $M \times M$. First a sub-sample $L$ of $M ^2$ size is taken from the raw signal $R$, hence, the $i$-th sub-sample is given by $L_i = \{R(i\cdot s+1), R(i\cdot s+2), ..., R(i\cdot s+M^2)\}$, where $s\in \mathbb{Z}^+$ is the step between samples, and the index $i=\{0,1,...,\lfloor N/s \rfloor \}$, with $\lfloor \ \rfloor$ denoting the floor function (see Fig. \ref{fig:signalToImage}). Note that we aim for an important overlap between samples in order to obtain more images, which is advantageous for the CNN classifier, i.e. $s<<M^2$. Then, each point in the sub-sample $L$ fills a matrix of $M \times M$ from left to right and from top to bottom. Each point is normalized from $0$ to $255$, and represents the grayscale intensity value of each pixel, with coordinates $x,y$, of the image $P(x, y)$. For the $i$-th image, this process is defined as follows

\begin{equation}
    P_i(x, y) = round\left ( \frac{L_i((x - 1) \cdot M + y) - min(L_i)}{max(L_i) - min(L_i)} \cdot 255 \right )
    \label{Ec:signalToImage}
\end{equation}
\vspace{3mm}

The size and number of images may vary according to the amount of vibration data available. In addition, the computational complexity will also be proportional to the size of the images. Henceforth, in case that complexity is a problem, the size of the images should be reduced \cite{Do2011}. For our proposal we have chosen a size of $64 \times 64$ pixels, with a step $s=64$.

\subsection{Convolutional Neural Network}

The CNN are deep neural networks that focus mainly on image processing and are excellent for pattern recognition. In addition, it is one of the best methods for classification. CNN automatically obtain the characteristics of the images by means of convolutional filters, which makes them a tool with a great capacity to learn characteristics in a robust and sensitive way.

In each CNN there are three main types of layers: a) the convolutional layer (Conv), b) the sub-sampling layer and c) the fully connected layer (FC). The convolutional layer serves to acquire feature maps that are obtained through a set of filters. The sub-sampling layer serves to reduce the characteristics of the inputs and the computational complexity. Finally, the FC layer, that is a layer of a normal neural network where each pixel is considered as a neuron, functions to calculate the scores of each of the classes \cite{Ren2017,Wen2018}. FC layers have a loss function as a SVM or softmax classifier \cite{Xie2017}.

\begin{figure*}[th!] %CNN AlexNet
\begin{center}
  \includegraphics[width=1\textwidth]{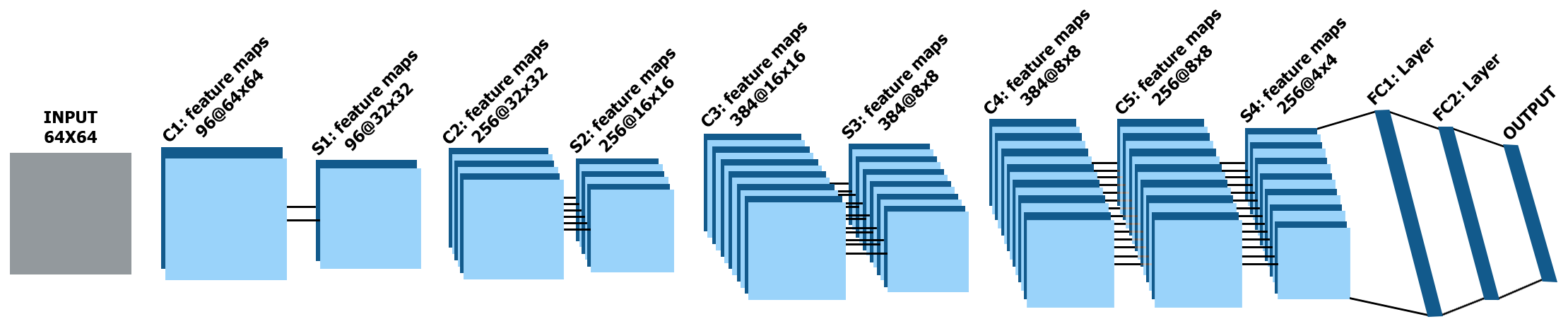}
\caption{Proposed architecture based on AlexNet for the estimation of bearing wear. The size of the kernels in the first and second convolutional layers was changed, also a sub-sampling layer of maxpooling was added between the third and fourth convolutional layers, finally the sub-sampling layers were resized and the FC layers changed the number of neurons with respect to the original AlexNet proposal.}
\label{fig:CNN AlexNet}       % Give a unique label
\end{center}
\end{figure*}

The CNN model we propose for the classification task is based on the state-of-the-art architecture AlexNet introduced by \cite{Krizhevsky1012}, and is shown in Fig. \ref{fig:CNN AlexNet}, while the configuration of the layers is presented in Table \ref{table:layersAlexNetProp}. From there, five convolutional layers (Conv), four sub-sampling layers with maxpooling (Maxpool) and three FC layers are applied, where the last one is the output layer. We can note that the sizes of the filters on each layer was modified with respect to the original proposal of AlexNet, in order to deal with smaller images. Once the signals have been converted into images, the CNN training can be performed to classify the level of wear of the bearings. 

%\textcolor{red}{no explicas en que consisten los layers, y no los defines nunca}

\begin{table}[hb!] %layerAlexNet
\centering
 \caption{Layers configuration for the proposed architecture, where $n$ indicates the number of neurons and $x$ the number of classes.}
 \begin{tabular}{c c}  
 \hline \\
 Layer & Features \vspace{0.2cm}\\
 \hline \\
 1 & $ Conv(5 \times 5 \times 96) $ \\
 2 & $ Maxpool(2 \times 2) $ \\
 3 & $ Conv(3 \times 3 \times 256)  $ \\
 4 & $ Maxpool(2 \times 2)  $ \\
 5 & $ Conv(3 \times 3 \times 384)  $ \\
 6 & $ Maxpool(2 \times 2) $ \\
 7 & $ Conv(3 \times 3 \times 384)  $ \\
 8 & $ Conv(3 \times 3 \times 256) $ \\
 9 & $ Maxpool(2 \times 2) $ \\
 10 & $ FC(n) $ \\
 11 & $ FC(n) $ \\
 12 & $ FC(x) $ \\  [1ex] 
 \hline
 \end{tabular}
\label{table:layersAlexNetProp}
\end{table}

The main modifications to the AlexNet architecture were made in order to deal with smaller images, resulting in a reduction of the kernel size used for each layer. Accordingly, in the first convolutional layer, the kernel size was reduced from $11 \times 11$ to $5 \times 5$; also the second convolutional layer kernel size was reduced from $5 \times 5$ to $3 \times 3$; while the maxpooling layers were reduced from $3 \times 3$ to $2 \times 2$. Furthermore, a new layer of maxpooling was added between the third and fourth convolutional layers. The use of smaller images is convenient since it allows us to obtain more images from available datasets, which is advantageous for the training algorithm. Furthermore, smaller images significantly reduce the computational cost for the training algorithms. Also, in the first two FC layers, the number of neurons was varied, looking for the best configuration for our particular case, using values from $512$ to $3584$ neurons, in the first FC layer, and from $0$ to $1024$ in the second one. Finally, the last layer only was changed according to the number of labeled classes, seven in our case.

\section{Experiments and results}
\label{Experiments and results}

The proposed methodology was implemented and tested with the University of Cincinnati's Center for Intelligent Maintenance Systems (IMS) \cite{NationalAeronauticsandSpaceAdministration2018} unlabeled dataset, where the results of each of the phases of the proposed methodology are shown in the following.

\subsection{Dataset and experimental setup}

\begin{figure}[th!] %experimentalSetup
\begin{center}
 \includegraphics[width=0.48\textwidth]{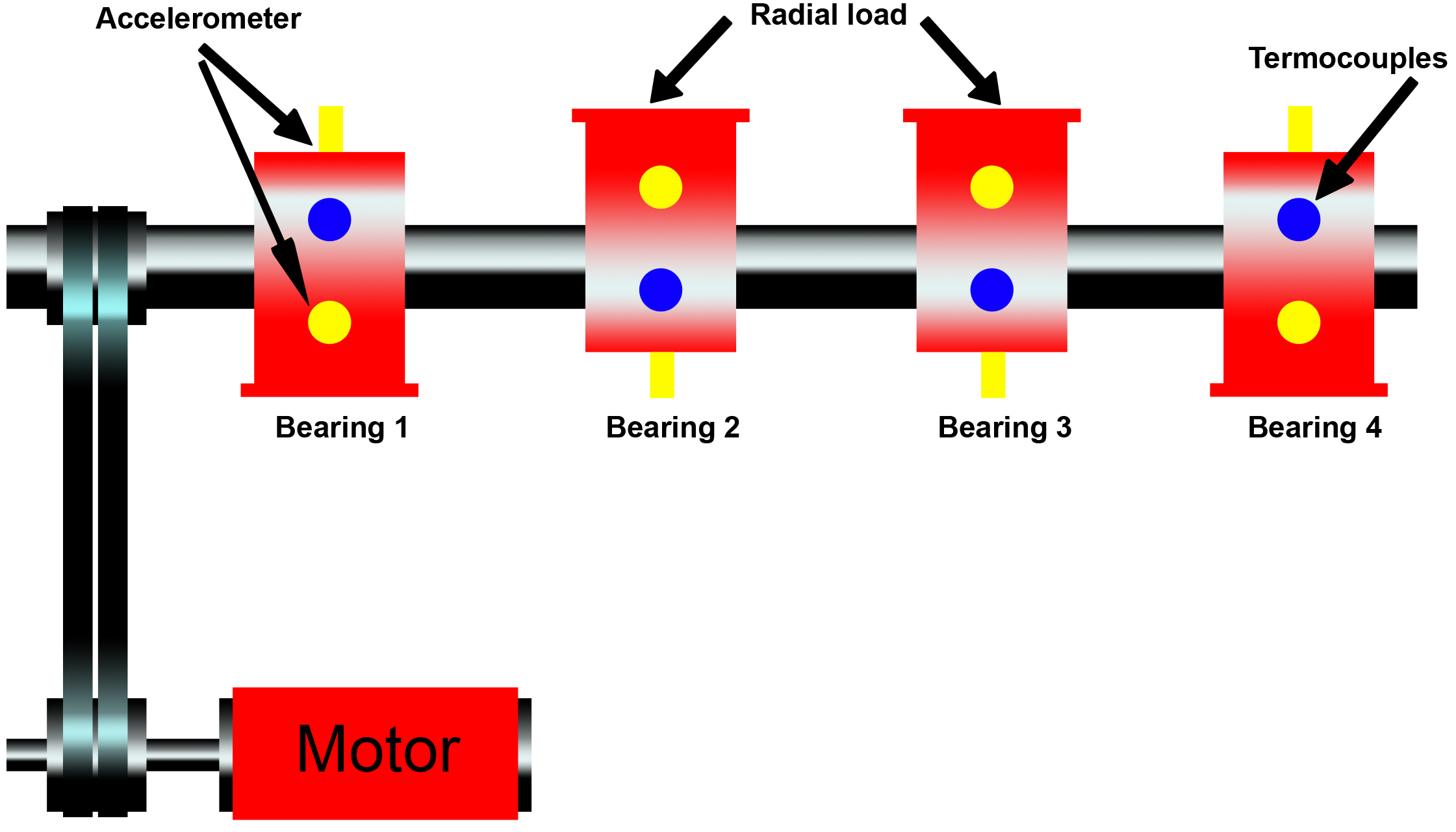}
\caption{Experimental setup for vibration data acquisition from bearings provided by \cite{NationalAeronauticsandSpaceAdministration2018}. It is conformed by an AC motor rotating at a constant speed of $2,000\ RPM$. Four bearings are mounted on the shaft, each one with two high-sensitivity accelerometers. Data is collected until failure, for one second, every 10 minutes, for a total of $20,480$ points. }
\label{fig:experimentalSetup}       % Give a unique label
\end{center}
\end{figure}

A vibration signal dataset is freely provided by the IMS \cite{NationalAeronauticsandSpaceAdministration2018}.
This dataset is one of the most used in the literature, and it was selected because it does not contain labels, allowing us to propose and validate a labeling method according to the level of wear. On the other hand, the bearings of this dataset were not physically altered to force failure, but the failures were presented in a natural way due to degradation in normal operation. As depicted in Fig. \ref{fig:experimentalSetup}, the experimental setup is composed of an AC motor running at a constant speed of $2,000\ RPM$, which is connected to the shaft by friction bands. Four Rexnord ZA-2115 double row bearings are mounted on the shaft, along with two high-sensitivity quartz accelerometers PCB 353B33 for each bearings. Moreover, a radial load of $6,000$ $lbs$ was applied to the shaft and bearing by means of a spring mechanism. The data was collected with the data acquisition system NI DAQCard 6062E. Failures occurred after the stress test exceeded the life time of the bearings. The experimental platform and the location of the sensors are shown in Fig. \ref{fig:experimentalSetup}.

The IMS dataset contains three failure tests, where the system is run under regular operation conditions until a failure occurs, produced by the deterioration of a different bearing each time. Each of the tests contains files, recording a snapshot of one second of the accelerometers vibration signal, which is stored in $10$ minutes time intervals. Each file has $20,480$ points with the sampling frequency at $20\ kHz$, including information of the eight accelerometers with a timestamp. Although this dataset is commonly used in the literature, it does not provides labels according to the wear level, hence it is not suitable for supervised training with CNN. This issue is overcome by the proposed automatic labeling method previously stated.

For the training, the CNNs were carried out in Python 3.6 with TensorFlow 1.12, and implemented on a computer equipped with a dedicated Graphic Processing Unit (GPU) NVIDIA GeForce RTX 2070, a processor Intel i7-9750H CPU and $16\ GB$ of RAM memory.

\subsection{Feature extraction}

\begin{figure*}[th!] %TSF
\begin{center}
  \includegraphics[width=0.9\textwidth]{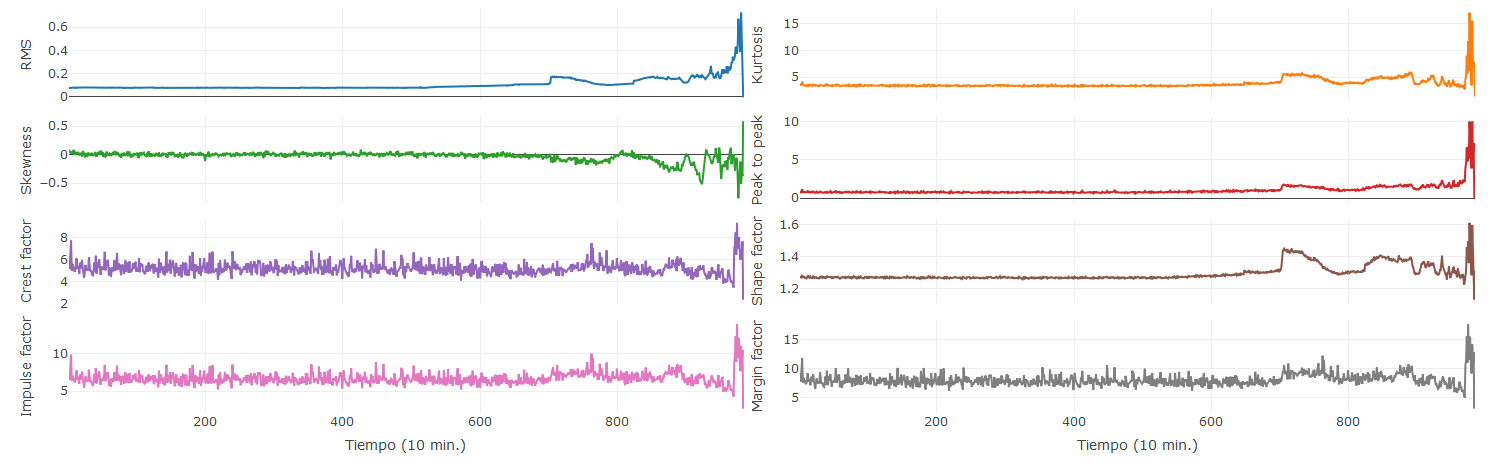}
\caption{ Traditional Statistical Features (TSF) for bearing 1. The results obtained with Crest factor, Impulse factor and Margin factor do not present an increasing or decreasing trend, which is undesired. While RMS, Kurtosis, Peak to peak and Shape factor present a more defined trend. }
\label{fig:TSF}       % Give a unique label
\end{center}
\end{figure*}

In order to identify the best feature extractor, all of the TSF in Table \ref{table:TraditionalStatisticFeatures} were applied to the accelerometer data provided by the IMS dataset, and the obtained plots are displayed in Fig. \ref{fig:TSF}. It was observed graphically that some measures may be discarded because there is not a well-defined increasing or decreasing trend, as is the case of Crest factor, Impulse factor and Margin factor. Also, it can be observed that RMS, Kurtosis, Peak to peak and Shape factor present a better defined increasing trend than the others, which helps to identify changes in vibration signals in the bearings, facilitating the classification.

\subsection{Shannon's entropy}

\begin{figure*}[th!] %entropy
\begin{center}
  \includegraphics[width=0.9\textwidth]{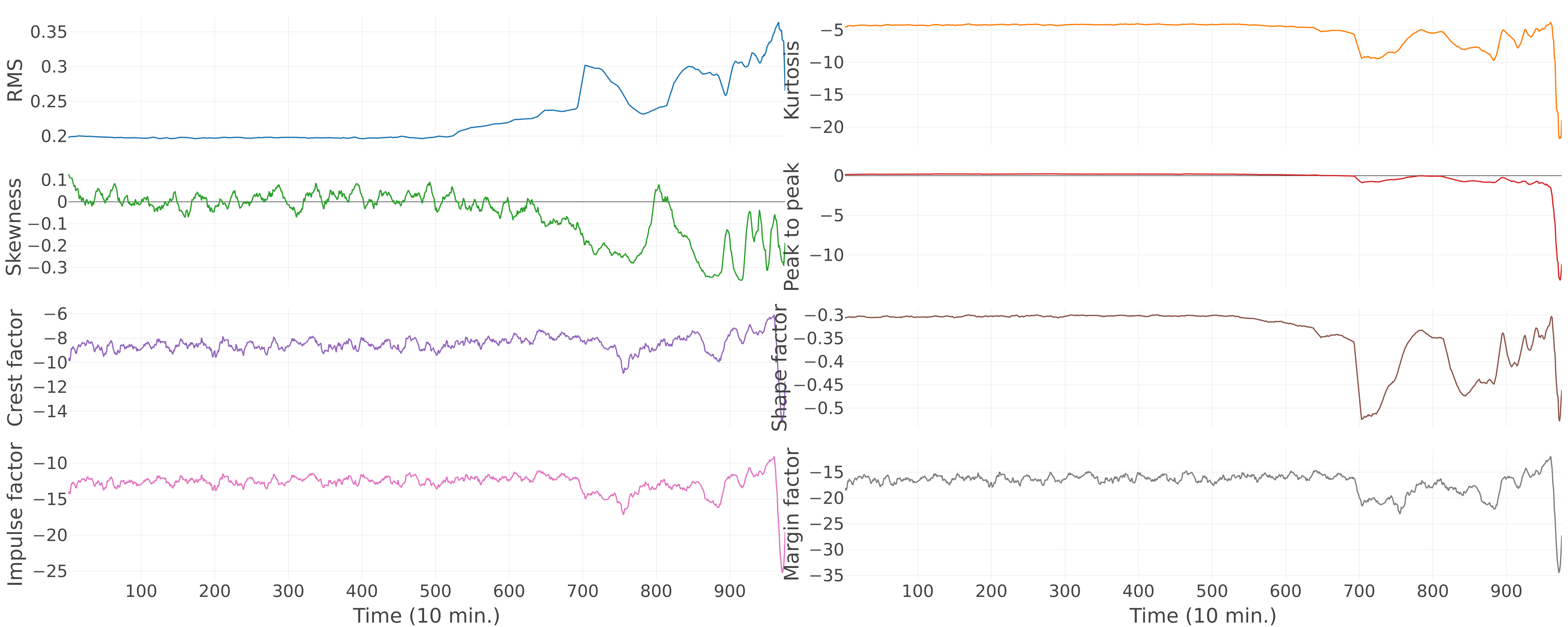}
\caption{Shannon's entropy measures of each for the TSF shown in Table \ref{table:TraditionalStatisticFeatures} for bearing 1. RMS is selected because it was the measure that grew the fastest over time, while showing variations from early wear stages.}
\label{fig:entropy}       % Give a unique label
\end{center}
\end{figure*}

Now, as shown in Fig. \ref{fig:entropy}, the Shannon's entropy is also calculated for each of the TSF, considerably stressing the effect of each feature, making it easier to identify the trends. Then, as mentioned before, the characteristics extracted from Crest factor, Impulse factor and Margin factor did not show a significant change, as well as Peak to peak, reason why they are discarded in this work. Skewness is also discarded because it decreases too slowly. We can observe that the TSF that show a more pronounced growth over time are RMS, Kurtosis and Shape factor, but we can see that Kurtosis does not show much variation at the beginning, that is, it is less sensitive to early wear in the bearings when compared to RMS or Shape factor. On the other hand, RMS and Shape factor present slight variations from the beginning, and are more suitable to characterize the level of wear from early stages. Henceforth, the combination of RMS with Shannon's entropy shows a suitable behavior to classify the level of wear on the bearings during their whole useful life, since it provides faster growth over time, which is essential to find the deterioration of the bearing, so it is selected for this methodology.

\subsection{Feature clustering}

\begin{figure}[th!] %entropy
\begin{center}
  \includegraphics[width=0.48\textwidth]{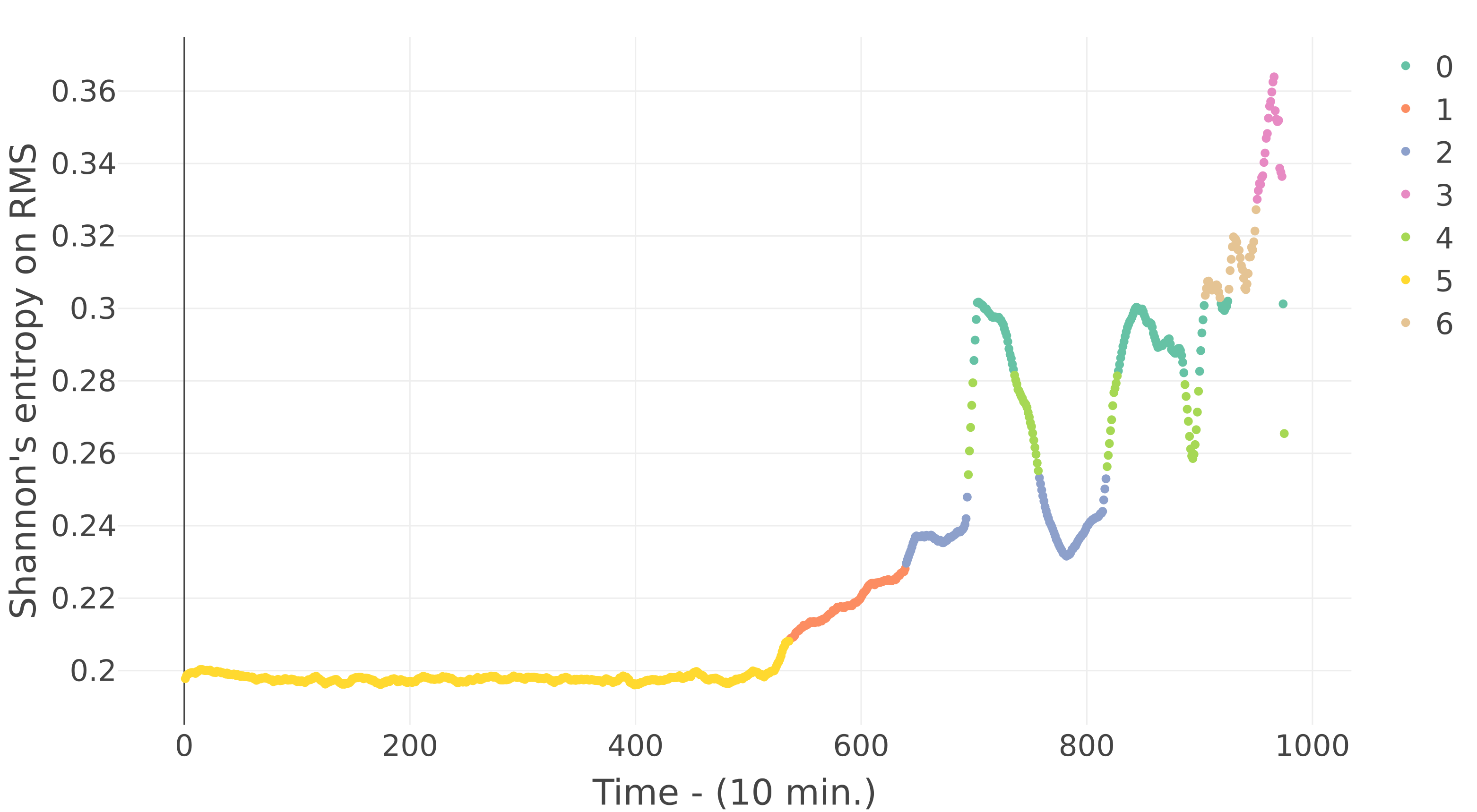}
\caption{Clusters made to separate the wear level of the bearings, using the K-means algorithm along with the Shannon's entropy and RMS.}
\label{fig:cluster}       % Give a unique label
\end{center}
\end{figure}

The data generated by the Shannon's entropy and RMS is then grouped to form seven clusters, where each cluster represents the level of wear that the bearing has over time. These clusters are considered as the classes to be identified by the CNN. Fig. \ref{fig:cluster} presents, with different colors, the classification obtained by the K-means algorithm applied to the Shannon's entropy with RMS signal, which is used to label the raw signals for the training algorithm. Similar to the classification employed by \cite{BenAli2015}, these classes were divided in the following class names according to the approximate level of wear: $0\% - 9\%$, $10\% - 24\%$, $25\% - 39\%$, $40\% - 54\%$, $55\% - 69\%$, $70\% - 84\%$ and $85\% - 100\%$.

This concludes the labeling method, which is of great relevance when dealing with unlabeled datasets, since it provides an automatic way to categorize the data, without the need of expert's supervision, or faking the bearing's wear by physically damaging them. This labeling can then be used for training a supervised learning classifier to estimate the level of wear. On the other hand, images can be generated from vibration signals with their corresponding label, from where we can implement classification models by means of CNN, as explained in the following. 

\subsection{Signal to image transformation}

\begin{figure} %imagesLabels
\centering
\begin{tabular}{c c c}
  \includegraphics[width=0.14\textwidth]{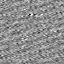} & \includegraphics[width=0.14\textwidth]{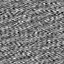} & \includegraphics[width=0.14\textwidth]{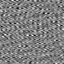}\\
 (a) $0\% - 9\%$ & (b) $10\% - 24\%$ & (c) $25\% - 39\%$ \\[6pt]
  \includegraphics[width=0.14\textwidth]{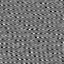} & \includegraphics[width=0.14\textwidth]{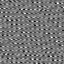}  & \includegraphics[width=0.14\textwidth]{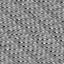}\\
 (d) $40\% - 54\%$ & (e) $55\% - 69\%$ & (f) $70\% - 84\%$ \\[6pt]
\multicolumn{3}{c}{\includegraphics[width=0.14\textwidth]{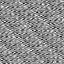} }\\
\multicolumn{3}{c}{(g) $85\% - 100\%$}
\end{tabular}
\caption{Conversion of accelerometer signals to images. The images represent the approximate wear of the bearings, it is (a) $0\% - 9\%$, (b) $10\% - 24\%$, (c) $25\% - 39\%$, (d) $40\% - 54\%$, (e) $55\% - 69\%$, (f) $70\% - 84\%$ and (g) $85\% - 100\%$ level of wear, respectively. }
\label{fig:imagesLabels}
\end{figure}

The result of the transformation of the raw signals into images is shown in Fig. \ref{fig:imagesLabels}, where each of the images represents a level of wear. The labeling of these levels was performed in the previous step, where seven different levels were obtained.

The total number of converted images was $251,904$ with a size of $64 \times 64$ pixels, using a step $s=64$. The use of a small step allows for an important overlap between sub-samples to form images, resulting in an increase in the number of images obtained, which is important to obtain a better performance of the CNN training. These images are labeled according to the classification obtained in the previous stage, and the number of images per cluster can be observed in Table \ref{table:numberOfImages}. The number of images for each of the classes according to the level of wear is unbalanced, due to the proportion of data that each of the clusters had, evidently, less information is available for the last wear levels near to failure. Performing a CNN training with the unbalanced data load in the classes, causes the CNN not to learn properly, resulting in erroneous estimates by the classifier. Therefore, it is very important that the images of each of the classes are balanced, that is, that each of the classes contains the same number of images. In this work, this is accomplished by generating a large amount of images, and randomly selecting the same number of images for each class.

\begin{table}[hb!] %numberOfImages
\centering
 \caption{Number of images per class.}
 \begin{tabular}{c c}  
 \hline \\
 Level & Number of images \vspace{0.2cm}\\
 \hline \\
 $0\% - 9\%$ & $139,520$   \\
 $10\% - 24\%$ & $26,368$  \\
 $25\% - 39\%$ & $29,440$   \\
 $40\% - 54\%$ & $13,056$   \\
 $55\% - 69\%$ & $27,648$  \\
 $70\% - 84\%$ & $9,984$   \\
 $85\% - 100\%$ & $5,888$  \\  [1ex] 
 \hline
 \end{tabular}
\label{table:numberOfImages}
\end{table}

\subsection{Convolutional Neural Network}
\begin{table}[b!] %layerWen2018
\centering
\caption{Layers configuration of the architecture proposed in \cite{Wen2018} based on LeNet-5, which is used for comparison, where $n$ indicates the number of neurons and $x$ the number of classes.}
 \begin{tabular}{c c}  
 \hline \\
 Layer & Features \vspace{0.2cm}\\
 \hline \\
 1 &  Conv($5 \times 5 \times 32$)  \\
 2 &  Maxpool($2 \times 2$) \\
 3 &  Conv($3 \times 3 \times 64$)  \\
 4 &  Maxpool($2 \times 2$)  \\
 5 &  Conv($3 \times 3 \times 128$)  \\
 6 &  Maxpool($2 \times 2$) \\
 7 &  Conv($3 \times 3 \times 256$) \\
 8 &  Maxpool($2 \times 2$) \\
 9 &  FC($n$) \\
 10 &  FC($n$) \\
 11 &  FC($x$) \\  [1ex] 
 \hline
 \end{tabular}
\label{table:layersWen2018}
\end{table}

\begin{table*}[h!] %AlexNet1FC
\centering
\caption{Results of the models of our AlexNet-based proposal with one FC layer. It can be seen that the model CNN-2560 obtained the best results in maximum, minimum and mean for each of the metrics.}
\begin{tabularx}{\textwidth}{X X X X X X X X}
\hline
 No. & CNN-512 & CNN-1024 & CNN-1536 & CNN-2048 & CNN-2560 & CNN-3072 & CNN-3584 \\
\hline
 \hline
 \multicolumn{8}{c}{Accuracy} \\
 \hline
 Max &  97.09\%  &  96.12\%  &  97.55\%  &  97.84\%  &  \textbf{97.96\%}  &  97.21\%  &  97.74\%  \\
 Min &  90.42\%  &  91.17\%  &  90.93\%  &  93.98\%  &  \textbf{95.27\%}  &  94.64\%  &  95.12\%  \\
 Mean &  94.59\%  &  94.61\%  &  94.74\%  &  96.03\%  &  \textbf{96.78\%}  &  96.36\%  &  96.19\%  \\
 Std &  2.34  &  1.4841  &  2.0931  &  1.1987  &  0.8178  &  \textbf{0.7689}  &  0.9575  \\ 
 \hline
 \multicolumn{8}{c}{Precision} \\
 \hline
 Max &  97.09\%  &  96.17\%  &  97.56\%  &  97.88\%  &  \textbf{98.00\%}  &  97.23\%  &  97.77\%  \\
 Min &  90.77\%  &  91.70\%  &  90.99\%  &  94.15\%  &  \textbf{95.34\%}  &  94.75\%  &  95.26\%  \\
 Mean &  94.78\%  &  94.77\%  &  94.84\%  &  96.08\%  &  \textbf{96.86\%}  &  96.42\%  &  96.28\%  \\
 Std &  2.2386  &  1.3458  &  2.0843  &  1.1704  &  0.8042  &  \textbf{0.7411}  &  0.9141  \\ 
 \hline
 \multicolumn{8}{c}{Recall} \\
 \hline
 Max &  97.09\%  &  96.12\%  &  97.55\%  &  97.84\%  &  \textbf{97.96\%}  &  97.21\%  &  97.74\%  \\
 Min &  90.42\%  &  91.17\%  &  90.93\%  &  93.98\%  &  \textbf{95.27\%}  &  94.64\%  &  95.12\%  \\
 Mean &  94.59\%  &  94.61\%  &  94.74\%  &  96.03\%  &  \textbf{96.78\%}  &  96.36\%  &  96.19\%  \\
 Std &  2.34  &  1.4841  &  2.0931  &  1.1987  &  0.8178  &  \textbf{0.7689}  &  0.9575  \\ 
 \hline
 \multicolumn{8}{c}{F1} \\
 \hline
 Max &  97.09\%  &  96.11\%  &  97.55\%  &  97.85\%  &  \textbf{97.97\%}  &  97.22\%  &  97.75\%  \\
 Min &  90.47\%  &  91.12\%  &  90.87\%  &  93.97\%  &  \textbf{95.29\%}  &  94.65\%  &  95.16\%  \\
 Mean &  94.60\%  &  94.60\%  &  94.74\%  &  96.03\%  &  \textbf{96.79\%}  &  96.36\%  &  96.19\%  \\
 Std &  2.3364  &  1.4944  &  2.1091  &  1.2064  &  0.8176  &  \textbf{0.7639}  &  0.9579  \\ 
 \hline
 \multicolumn{8}{c}{MSE} \\
 \hline
 Max &  0.09993  &  0.08974  &  0.09144  &  0.06233  &  \textbf{0.04730}  &  0.05506  &  0.04948  \\
 Min &  0.02983  &  0.03978  &  0.02522  &  0.02231  &  \textbf{0.02159}  &  0.02862  &  0.02328  \\
 Mean &  0.05547  &  0.05559  &  0.05384  &  0.04097  &  \textbf{0.03301}  &  0.03730  &  0.03888  \\
 Std &  0.0243  &  0.0151  &  0.0213  &  0.0124  &  0.0079  &  \textbf{0.0077}  &  0.0096  \\ 
\hline
\end{tabularx}
\label{table:AlexNet1FC}
\end{table*}

\begin{table*}[h!] %AlexNet2FC
\centering
\caption{Results of the models of our AlexNet-based proposal with two FC layers. It can be seen that the model CNN-2560-256 obtained the best results in maximum, minimum and mean for the metrics accuracy, precision, recall, F1.}
\begin{tabularx}{\textwidth}{X X X X X X X X X}
\hline
 No. & CNN-2560 & CNN-2560-64 & CNN-2560-128 & CNN-2560-256 & CNN-2560-512 & CNN-2560-768 & CNN-2560-1024 & LeNet5-2560-512 \\
\hline
 \hline
 \multicolumn{9}{c}{Accuracy} \\
 \hline
 Max &  97.96\%  &  98.30\%  &  98.13\%  &  \textbf{99.25\%}  &  97.67\%  &  97.11\%  &  97.60\% & 97.04\%  \\
 Min &  95.27\%  &  94.18\%  &  93.82\%  &  \textbf{95.54\%}  &  92.58\%  &  93.28\%  &  94.71\% &  93.50\% \\
 Mean &  96.78\%  &  96.19\%  &  95.78\%  &  \textbf{96.84\%}  &  95.33\%  &  95.57\%  &  96.29\% & 95.38\%  \\
 Std &  \textbf{0.8178}  &  1.1948  &  1.3116  &  1.0581  &  1.5948  &  1.1762  &  0.889 & 1.1045  \\ 
 \hline
 \multicolumn{9}{c}{Precision} \\
 \hline
 Max &  98.00\%  &  98.32\%  &  98.16\%  &  \textbf{99.25\%}  &  97.72\%  &  97.20\%  &  97.63\% & 97.07\%  \\
 Min &  95.34\%  &  94.35\%  &  94.04\%  &  \textbf{95.56\%}  &  92.89\%  &  93.55\%  &  94.79\% & 93.83\%  \\
 Mean &  96.86\%  &  96.27\%  &  95.89\%  &  \textbf{96.90\%}  &  95.50\%  &  95.73\%  &  96.36\% & 95.49\%  \\
 Std &  \textbf{0.8042}  &  1.136  &  1.2596  &  1.0352  &  1.5031  &  1.119  &  0.9012 & 1.0505  \\ 
 \hline
 \multicolumn{9}{c}{Recall} \\
 \hline
 Max &  97.96\%  &  98.30\%  &  98.13\%  &  \textbf{99.25\%}  &  97.67\%  &  97.11\%  &  97.60\% & 97.04\%  \\
 Min &  95.27\%  &  94.18\%  &  93.82\%  &  \textbf{95.54\%}  &  92.58\%  &  93.28\%  &  94.71\% & 93.50\%  \\
 Mean &  96.78\%  &  96.19\%  &  95.78\%  &  \textbf{96.84\%}  &  95.33\%  &  95.57\%  &  96.29\% & 95.38\%  \\
 Std &  \textbf{0.8178}  &  1.1948  &  1.3116  &  1.0581  &  1.5948  &  1.1762  &  0.889 & 1.1045  \\ 
 \hline
 \multicolumn{9}{c}{F1} \\
 \hline
 Max &  97.97\%  &  98.31\%  &  98.14\%  &  \textbf{99.25\%}  &  97.67\%  &  97.11\%  &  97.61\% & 97.05\%  \\
 Min &  95.29\%  &  94.20\%  &  93.83\%  &  \textbf{95.53\%}  &  92.56\%  &  93.26\%  &  94.72\% & 93.55\%  \\
 Mean &  96.79\%  &  96.20\%  &  95.78\%  &  \textbf{96.84\%}  &  95.33\%  &  95.57\%  &  96.30\% & 95.39\%  \\
 Std &  \textbf{0.8176}  &  1.1932  &  1.316  &  1.0608  &  1.6058  &  1.1809  &  0.89 & 1.0945  \\ 
 \hline
 \multicolumn{9}{c}{MSE} \\
 \hline
 Max &  \textbf{0.04730}  &  0.05894  &  0.06258  &  0.05336  &  0.08537  &  0.07082  &  0.05700 & 0.06718  \\
 Min &  0.02159  &  0.01771  &  0.01940  &  \textbf{0.00825}  &  0.02474  &  0.02959  &  0.02474 & 0.03032  \\
 Mean &  \textbf{0.03301}  &  0.03936  &  0.04312  &  0.03374  &  0.04994  &  0.04625  &  0.04046 & 0.04749  \\
 Std &  \textbf{0.0079}  &  0.0124  &  0.0132  &  0.0123  &  0.0187  &  0.0126  &  0.0109 & 0.0115  \\ 
\hline
\end{tabularx}
\label{table:AlexNet2FC}
\end{table*}

In order to obtain a balanced number of images for each class, an equal amount of images is selected randomly. This way, the loading of each of the classes was done in a balanced way for the CNN to learn equally. Therein the importance of counting with a large number of images, provided that less data is available about the last stages of the bearing wear before failure. Moreover, the images were used $70\%$ for training and $30\%$ for testing.

Furthermore, different number of neurons were tested in the FC, convolutional and sub-sampling layers, to find the best configuration for this task. Each one of the CNN combinations was executed ten times, where the maximum, minimum, average and standard deviation of each of the acquired metrics were obtained. The metrics considered in each CNN training are: accuracy, precision, recall, F1 and the Mean Square Error (MSE) \cite{Alla2019}. 

The proposed CNN was run with one and two FC layers, trying different number of neurons, looking for the best configuration. The name of each model is denoted by the form CNN-$i$-$j$, where the values of $i$ and $j$ represent the number of neurons in the corresponding FC layers, one and two, respectively. For example, CNN-512 represents that the first FC layer has 512 neurons and the second FC layer does not have any. For CNN-1024-64 means that the first FC layer has 1024 neurons and the second FC layer has 64 neurons.

The first experiment was done with the proposed CNN with a single FC layer, with the purpose of finding the adequate number of neurons while disregarding the second FC layer. The number of neurons was changed between $512$, $1024$, $1536$, $2048$, $2560$, $3072$ and $3584$, resulting in seven different models according to the number of neurons used, as shown in Table \ref{table:AlexNet1FC}. From there, it can be observed that with the model CNN-2560 we obtained the best accuracy, precision, recall and F1 of $97.96\%$, $98\%$, $97.96\%$ and $97.97\%$ respectively. Also, we obtained the lowest error of $0.02159$, as well as the minimum and mean of all the metrics. With the model CNN-3072 results were lower than those of the model CNN-2560, but the variability between results was lower with CNN-3072. It can be observed that in the proposed CNN experiment with only one FC layer with 2560 neurons (CNN-2560), good results can be obtained, when compared with the rest of the configurations, hence it was selected for the next step.

Once the best number of neurons for the first FC layer is obtained, the second FC layer can be added. Then, the next experiment is to find the best number of neurons for the second FC layer, where the best first layer obtained from the previous step is employed, it is using 2560 neurons, and the second FC layer is tested with values of $64$, $128$, $256$, $512$, $768$ and $1024$ neurons. Then, six models of the proposed CNN with two FC layers are tested based on the CNN-2560 model, where the results can be seen in Table \ref{table:AlexNet2FC}. The maximum percentages of the results for accuracy, precision, recall and F1 were $99.25\%$, and the minimum error of $0.00825$ was obtained with the model CNN-2560-256. We can also appreciate the effect of adding the second FC layer when compared with the best result using only one FC, resulting in an improvement when selecting the appropriate number of neurons.

In addition to the proposed architecture, other architectures were implemented to compare the results. In \cite{Wen2018}, an architecture for classifying bearing failures was proposed based on LeNet-5, which is one of the best reported in the literature, hence it is used for comparison with our experiments. The configuration of the layers of this architecture is shown in Table \ref{table:layersWen2018}. In order to provide a fair comparison, the procedure in \cite{Wen2018} was implemented for the IMS dataset, and different numbers of neurons were also tested in the two first FC layers, in the same fashion as for our proposal, finding out that the best results for the experiment with a single FC layer were achieved with 2560 neurons in the first FC layer. For the second experiment the second FC layer was added, and the best results were achieved with 512 neurons in the second FC layer, that is, with the LeNet5-2560-512 model. The obtained results for this model are also included in the last column of Table \ref{table:AlexNet2FC}, where we can appreciate the superior performance of our proposal with respect to this LeNet5-based architecture.

On the other side, to further evaluate the results of this work, the original proposals of LeNet-5 and AlexNet architectures were also tested and compared in our study, along with other CNN and classical methods, such as SVM and ANN, among others. The comparison results are shown in the Table \ref{table:otherExperiments}. We can appreciate that the original proposal of AlexNet is not well suited for this problem before adaptation, mainly due to the large size of the filters in the convolutional layers. Using traditional methods, such as ANN and SVM, the results were very low to be used as classifiers. In addition, training with SVM is very time consuming, even with the reduced number of features. With diffuse learning, using the SFAM method, the results reported are very low. One dimensional convolutional neural networks such as the DCTLN and 1DCNN methods are only able to find the type of failure that occurs in the bearing, but they do not classify the wear that the bearings have over time, and the results are not very promising. We can observe that our proposal provides excellent results, clearly surpassing other methods reported in the literature for the IMS dataset.

In summary, RMS in conjunction with Shannon's entropy proved to be an excellent option for feature extraction, because it detects changes in vibration signals over time more quickly. Furthermore, the transformation of vibration signals into images provides a good way to analyze features in two dimensions, and along with CNN are excellent for pattern recognition, performing feature extraction automatically and learning features robustly. On the other hand, after extensive experiments, the proposed CNN architecture based on AlexNet with two FC layers obtained the highest results in the present study with the IMS dataset, but specifically the CNN-2560-256 was the best model, significantly overcoming other techniques reported in the state-of-the-art literature.

\begin{table}[b!] %otherExperiments
\centering
 \caption{Comparison results with others methods.}
 \begin{tabular}{c c c}  
 \hline \\
 Rank & Methods & Accuracy \vspace{0.2cm}\\ 
 \hline \\
   1 & \textbf{Proposal} &  \textbf{ $99.25\%$ } \\
   2 & CEEMD \cite{Lu2020} & $98.50\%$ \\ 
   3 & DNN \cite{Zhang2017a} & $98.35\%$ \\
   4 & Based on LeNet-5 \cite{Wen2018} & $97.04\%$ \\
   5 & 1D CNN  \cite{Eren2019} & $93.90\%$ \\
   6 & LeNet-5 & $92.04\%$ \\
   7 & DCTLN \cite{Guo2019} & $86.30\%$ \\ 
   8 & SVM & $81.00\%$ \\ 
   9 & SFAM \cite{BenAli2015} & $65.46\%$ \\
   10 & AlexNet & $14.29\%$ \\ 
   11 & ANN & $14.14\%$ \\ [1ex] 
 \hline
 \end{tabular}
\label{table:otherExperiments}
\end{table}

\section{Conclusion and future work}
\label{Conclusion}

In this article, we present a method for estimating the level of bearing's wear, by vibrations analysis using images and CNN classifiers. This is important for industries that operate rotatory machinery, preventing failures that may result in stopping production, complete system failure, damaging expensive components or even accidents, hence avoiding significant economic losses. 

The proposed CNN architecture is based on AlexNet, and was extensively validated with the IMS dataset, obtaining an accuracy of $99.25\%$, which represents an important improvement with respect to previous results in the literature, including both classic techniques and state-of-the-art CNN based methods. 

This proposal is ideal to be used for unlabeled datasets, or new unclassified data. Accordingly, we proposed a technique to automatically make the labeling of unclassified datasets, without the supervision of an expert, or faking the bearings wear by physically damaging them. The proposed labeling strategy is accomplished by means of Root Mean Square (RMS) combined with the Shannon's entropy for feature extraction, and the K-means algorithm for unsupervised classification.

We have found that the use of small size images along with an important overlap between them is suitable for this kind of task, due to the limited amount of data available, since it allows to obtain a good amount of images, which is key for a good training with balanced classes. Henceforth, the AlexNet architecture was adapted to deal with small size images.

There are a few things that remain to be proven in future works. Particularly, we are interested in replicate the results with other datasets that exist in the literature, in order to further validate our proposal and compare it with other works. On the other side, we would like to try different architectures, such as VGG, and make modifications to adapt them to the problem under consideration.

\begin{acknowledgements}
This work was supported by the Mexican National Council of Science and Technology CONACYT, and the FORDECyT project 296737 “Consorcio en Inteligencia Artificial”.

\end{acknowledgements}

 \section*{Conflict of interest}
 The authors declare that they have no conflict of interest.

%\clearpage


\bibliographystyle{IEEEtran}
\bibliography{bibilography}

%\begin{thebibliography}{}
%\bibitem{Pool2017}

% etc
%\end{thebibliography}

\end{document}